\documentstyle[preprint,aps,epsf]{revtex}

\begin{document}
\draft
\preprint{KANAZAWA 99-09, June 1999}
\title{A Quantum Perfect Lattice Action for Monopoles and Strings }
\author{Shouji Fujimoto, Seikou Kato and Tsuneo Suzuki}
\address{Institute for Theoretical Physics, Kanazawa University, 
 Kanazawa 920-1192, Japan}
\date{\today}
\maketitle
\begin{abstract}
A quantum perfect lattice action in four dimensions 
can be derived analytically 
as a renormalized trajectory when 
we perform a block spin transformation of monopole currents in a
 simple but non-trivial case of quadratic monopole interactions.
The spectrum of the lattice theory is identical to that of the continuum 
 theory.
The perfect monopole action is transformed exactly into a lattice
 action of a string model. 
A perfect operator evaluating a static potential between electric 
charges is also derived explicitly. 
If the monopole interactions are weak as in the case of 
infrared $SU(2)$ QCD, the string interactions become strong. 
The static potential and the
 string tension is  estimated analytically by the use of the 
strong coupling expansion and 
the continuum rotational invariance is restored completely.

\vspace{3mm}
$PACS$: 12.38.Gc, 11.15.Ha\\
$Keywords$: lattice QCD; perfect action;block spin transformation;
 monopole; dual transformation; string model
\end{abstract}

\newpage
To obtain the continuum limit is crucial in the framework of lattice 
field theories. A block spin transformation which is one of
renormalization group transformations is adopted usually on the lattice.
A quantum perfect lattice action is an action 
on the renormalized trajectory on which one can take the 
continuum limit. 

In principle, we 
obtain the renormalized trajectory when we perform infinite steps 
(the $n\to\infty$ limit) of 
block spin transformations for fixed physical length $b=na$ where $a$ is 
the lattice constant. But this is actually impossible in ordinary cases.
What we can do in actual simulations is to approach the renormalized
trajectory carrying out as many steps of block spin transformations 
as possible on a finite lattice $N^4$. 
If the effective action $S(n,a,N)$ obtained
 satisfies well the two conditions that (1) $S(n,a,N)$ is a function 
of $b=na$ alone (the scaling behavior and volume independence) and 
(2) the continuum rotational
invariance is satisfied, then the 
effective action  could be 
regarded as a good approximation of the renormalized trajectory. 
The first condition can be checked when we compare $S(n,a,N)$ themselves 
for various $a$, $n$ and $N^4$. But to test the rotational invariance,
we have to determine the correct form of physical operators  
on the blocked lattice. It is the perfect lattice operator 
on the renormalized trajectory which reproduces the continuum rotational
invariance. To find the perfect lattice operator 
is highly nontrivial, too.

The purpose of this note is to give a simple but a non-trivial 
lattice model composed of general monopole quadratic interactions alone 
with which
a block spin transformation can be done analytically. The renormalized
trajectory and the perfect operator corresponding to a potential  
between static electric charges can be derived analytically.
This is similar to the blocking from the continuum theory as developed
by Bietenholz and Wiese\cite{wiese96}.
The spectrum of the lattice theory is the same as in the continuum theory.
The continuum rotational invariance is shown exactly with the operator.
In addition, this model is very interesting, since the effective
monopole action obtained after an abelian projection  
of pure $SU(2)$ lattice QCD is known to be well dominated 
by such quadratic monopole interactions alone in the infrared region
\cite{shiba2,shiba3,kato97,kato98}. 

Let us start from the following action composed of quadratic 
interactions between magnetic monopole currents.  
It is formulated on an infinite lattice 
with very small lattice constant $a$:
\begin{eqnarray}
S[k]=\sum_{s,s',\mu}k_\mu(s)D_0(s-s')k_\mu(s').\label{eqn.S}
\end{eqnarray}
Since we are starting from the region very near to the continuum limit,
it is natural to assume the direction independence of $D_0(s-s')$.
Also we have adopted only parallel interactions, since we can avoid 
perpendicular interactions from  
short-distance 
terms
using the current conservation.
Moreover, for simplicity, we adopt only the first three Laurent
expansions, i.e., Coulomb, self and nearest-neighbor 
interactions. Explicitly, $D_0(s-s')$ is expressed as 
$\beta\Delta_L^{-1}(s-s')+\alpha\delta_{s,s'}+\gamma\Delta_L(s-s')$.
Here $\Delta_L(s-s')=-\sum_\mu\partial_\mu\partial'_\mu\delta_{s,s'}$ 
and $\partial(\partial')$ is the forward (backward) difference.  
Including more complicated quadratic interactions is not difficult.

How to evaluate a static potential between electrically charged
particles is a problem.  
It is known that the theory with the above action (\ref{eqn.S})
 is equivalent to 
an abelian gauge theory of the Villain form\cite{smit,misha1}.
In this model, it is natural to use 
an abelian Wilson loop 
$W({\cal C})=\exp i\sum_{\cal C}(\theta_{\mu}(s),J_{\mu}(s))$,
where $\theta_{\mu}(s)$ is an abelian angle variable of the 
modified Villain action.
Also the theory with the above action (\ref{eqn.S}) 
can be rewritten\cite{misha1}
in the lattice form of the modified London limit of the dual 
abelian Higgs model\cite{suzu89}. 
The static potential is evaluated by a 'tHooft operator in the 
model. However the expectation values of both operators are not
completely equivalent, although the term of the area law is the same
\cite{kato9910}. 
When use is made 
of BKT transformation\cite{Bere,KT,misha1,kato9910}, we see that 
the area law term is given correctly also by 
the following operator in the monopole
action \footnote{Using the monopole definition
$\grave{\mbox{a}}$ la DeGrand-Toussaint, we can prove it also directly from
abelian Wilson loops\cite{stack,shiba1}.}:
\begin{eqnarray}
W_m({\cal C})&=&
\exp\bigg( 2\pi i\sum_{s,\mu}N_{\mu}(s,S^J)k_{\mu}(s) \bigg),
\label{eqn.WC}\\
N_{\mu}(s,S_J)&=&\sum_{s'}\Delta_L^{-1}(s-s')\frac{1}{2}
\epsilon_{\mu\alpha\beta\gamma}\partial_{\alpha}
S^J_{\beta\gamma}(s'+\hat{\mu}), \label{eqn.N}
\end{eqnarray}
where $S^J_{\beta\gamma}(s'+\hat{\mu})$ is a plaquette variable satisfying 
$\partial'_{\beta}S^J_{\beta\gamma}(s)=J_{\gamma}(s)$ and the coordinate 
displacement $\hat{\mu}$ is due to the interaction between dual
variables. 
It is possible to prove that any choice of $S^J_{\beta\gamma}(s)$ for
fixed electric currents $J_{\gamma}(s)$ gives the same value in the
continuum limit $a\to 0$, since the difference is a closed surface and 
the exponent of $W_m(C)$ for the difference is just the four-dimensional 
linking number times $2\pi i$. Hence we take a flat surface for 
$S^J_{\beta\gamma}(s)$ in the following. 
Since the area law term is the same, let us consider only the operator
 (\ref{eqn.WC}) in the following.
The details of the definition
of the operator evaluating the static potential are discussed in 
Ref.\cite{kato9910}.

Now let us define  
a blocked monopole current\cite{ivanenko}:
\begin{eqnarray}
K_\mu(s^{(n)})&=&
  \sum_{i,j,l=0}^{n-1}
    k_\mu(ns^{(n)}+(n-1)\hat{\mu}+i\hat{\nu}+j\hat{\rho}+l\hat{\sigma})
\nonumber \\
&\equiv&{\cal B}_{k_\mu}(s^{(n)}).
\end{eqnarray}
With this definition, the current $K_\mu(s^{(n)})$ 
on the coarse lattice with 
a lattice distance $b=na$ satisfies the current conservation 
$\partial'_{\mu}K_\mu(s^{(n)})=
\sum_{\mu}(K_{\mu}(s^{(n)})-K_{\mu}(s^{(n)}-b\hat{\mu}))=0$.
The block spin transformation is expressed as 
\begin{eqnarray}
Z[K,J] &=&
  \sum_{k_{\mu}=-\infty\atop{\partial^{\prime}_{\mu}k_{\mu}=0}}^{\infty}
  \exp
  \Bigg\{-\sum_{s,s',\mu} k_{\mu}(s)D_0(s-s')k_{\mu}(s')
+2\pi i\sum_{s,\mu} N_{\mu}(s)k_{\mu}(s)\Bigg\}\nonumber \\
&&\times
  \delta
      \bigg(K_{\mu}(s^{(n)})
       -{\cal B}_{k_\mu}(s^{(n)})
      \bigg).
\end{eqnarray}
The vacuum expectation value of the Wilson loop (\ref{eqn.WC}) is  
written in terms of 
$K_\mu(s^{(n)})$ as follows:
\begin{eqnarray}
\langle W_m({\cal C}) \rangle = 
  \sum_{K_{\mu}=-\infty\atop{\partial^{\prime}_{\mu}K_{\mu}=0}}^{\infty}
  \!\!\!\!
  Z[K,J]
  \Bigg/
  \sum_{K_{\mu}=-\infty\atop{\partial^{\prime}_{\mu}K_{\mu}=0}}^{\infty}
  \!\!\!\!
  Z[K,0].
\end{eqnarray}

Introducing auxiliary field $\phi$ and $\gamma$, we rewrite the constraints
$\partial^{\prime}_{\mu}k_{\mu}=0$ and
$K_{\mu}(s^{(n)})={\cal B}_{k_\mu}(s^{(n)})$.
Then we change the integral region of $\gamma$ and $\phi$
from the first Brillouin zone to the infinite region, since the monopole
currents take integer values.
Making use of the Poisson sum rule and recovering dimensional lattice
constants $a$ and $b=na$, we get
\begin{eqnarray}
Z[K,J] \; &\propto&
\int_{-\infty}^{\infty}{\cal D}\!\!\  \gamma
  \exp\Bigg\{
    ib^4 \sum_{x^{(n)},\mu} \gamma_{\mu}(bs^{(n)})K_{\mu}(bs^{(n)})
  \Bigg\}
  \nonumber\\
&&\times
\sum_{al=-\infty\atop{al \in Z}}^{\infty}
\int_{-\infty}^{\infty}\!\!{\cal D}\phi
\int_{-\infty}^{\infty}\!\!{\cal D}F
\;
\exp
\Biggl\{
  -a^8\!\!\sum_{s,s',\mu}\!\! F_{\mu}(as)D_0(as-as')F_{\mu}(as')
\nonumber \\
&&
  +ia^4\sum_{s,\mu}
  \Bigl[
    2\pi N_{\mu}(as)
    + \phi(as)\partial'_{\mu}
    + 2\pi l_{\mu}(as)
  \Bigr]F_{\mu}(as)
  -a^4\!\!\sum_{s^{(n)},\mu}\!\! n
    \gamma_{\mu}(nas^{(n)})
      {\cal B}_{F_\mu}(s^{(n)})
\Biggr\}.\nonumber \\
\end{eqnarray}
Since we take the continuum limit $a\to 0$ finally, $l=0$ alone may
remains in the sum with respect to $al \in Z$.  
Carrying out explicitly integrals with respect to $F$, $\phi$, $\gamma$
and taking the continuum limit $a\to 0$, we obtain the expectation
value of the operator and the effective action on the coarse lattice:
\begin{eqnarray}
\langle W({\cal C}) \rangle
&=&
  \exp\Biggl\{
    - \pi^2 \int_{-\infty}^{\infty}\!\! d^4xd^4y
    \sum_{\mu}N_{\mu}(x)D_0^{-1}(x-y)N_{\mu}(y)
\nonumber \\
&&
    + \pi^2 b^8\!\!\!\! \sum_{s^{(n)},s^{(n)'}\atop{\mu,\nu}}
    \!\!\!\!
    B_{\mu}(bs^{(n)})
      A_{\mu\nu}^{\prime{\rm GF} -1}(bs^{(n)}-bs^{(n)'})
    B_{\nu}(bs^{(n)'})
  \Biggr\}
\nonumber \\
&&
\times \!\!\!\!
\sum_{b^3K_\mu(bs)=-\infty\atop\partial'_\mu K_\mu=0}^{\infty}
\!\!\!\!\!\!
  \exp\Biggl\{
    - b^8\!\!\!\! \sum_{s^{(n)},s^{(n)'}\atop{\mu,\nu}}
    \!\!\!\!
    K_{\mu}(bs^{(n)})
      A_{\mu\nu}^{\prime{\rm GF} -1}(bs^{(n)}-bs^{(n)'})
    K_{\nu}(bs^{(n)'})
\nonumber \\
&&
    +2 \pi i b^8\!\!\!\! \sum_{s^{(n)},s^{(n)'}\atop{\mu,\nu}}
    \!\!\!\!
     B_{\mu}(bs^{(n)})
       A_{\mu\nu}^{\prime{\rm GF} -1}(bs^{(n)}-bs^{(n)'})
     K_{\nu}(bs^{(n)'})
  \Biggr\}
  \Bigg/
  \!\!\!\!
  \sum_{b^3K_\mu(bs)=-\infty\atop\partial'_\mu K_\mu=0}^{\infty}
  \!\!\!\!\!\! Z[K,0],
\label{opwil:1}
\end{eqnarray}
where
\begin{eqnarray}
B_\mu(bs^{(n)}) &\equiv&
\lim_{a\to 0 \atop{n\to\infty}}
  a^8\sum_{s,s',\nu}
    \Pi_{{\neg}\mu}(bs^{(n)}-as)
A_{\mu\nu}(as-as')N_{\nu}(as'),
\label{opwil:9}
\\
\Pi_{\neg\mu}(bs^{n}-as)&\equiv&
\frac{1}{n^3}
  \delta\left( nas_\mu^{(n)}+(n-1)a-as_\mu \right)
  \times
  \prod_{i(\ne \mu)}\left(
    \sum_{I=0}^{n-1}\delta\left( nas_i^{(n)}+Ia-as_i \right)
  \right),\\
A_{\mu\nu}(as-as')&\equiv&
\left\{
  \delta_{\mu\nu}
  -\frac{\partial_{\mu}\partial'_{\nu}}{\sum_{\rho}\partial_{\rho}
  \partial'_{\rho}}
\right\}D_0^{-1}(as-as').
\end{eqnarray}
$A_{\mu\nu}^{\prime{\rm GF} -1}(bs^{(n)}-bs^{(n)'})$ is a
gauge-fixed inverse of the following operator:
\begin{eqnarray}
A'_{\mu\nu}(bs^{(n)}-bs^{(n)'})\equiv
a^8\sum_{s,s'}
\Pi_{\neg\mu}(bs^{(n)}\!\!-as)
  A_{\mu\nu}(as-as')
\Pi_{\neg\nu}(bs^{(n)'}\!\!\!-as').
\label{amunupr}
\end{eqnarray}
Here we have used $\sum_\mu \partial_\mu N_\mu(s)=0$ and have adopted
a gauge including $\lambda\{\partial_\mu \gamma_\mu(bs^{(n)})\}^2$ in
the integral with respect to $\gamma$.

The momentum representation of the gauge fixed propagator is given
explicitly by
\begin{eqnarray}
A_{\mu\nu}^{\prime\rm GF}(p)=
\Biggl\{
  A'_{\mu\nu}(p)
  +\lambda\hat{p}_\mu\hat{p}_\nu
\Biggr\}
e^{i(p_\mu-p_\nu)/2},
\end{eqnarray}
where $\hat{p}_\mu=2\sin(p_\mu/2)$ and
$A'_{\mu\nu}(p)$ is written as
\begin{eqnarray}
A'_{\mu\nu}(p)\!&\equiv&\!
\left(\prod_{i=1}^4\sum_{l_i=-\infty}^{\infty} \right)
\!\!
\Biggl\{
  \!D_0^{-1}(p+2\pi l)
  \Biggl[
    \delta_{\mu\nu}\!-\frac{(p+2\pi l)_\mu(p+2\pi l)_\nu}{\sum_i(p
+2\pi l)_i^2}
  \Biggr]
  \frac{(p+2\pi l)_\mu(p+2\pi l)_\nu}{\prod_i(p+2\pi l)_i^2}
\Biggr\}\nonumber \\
&&\hspace{2cm}\times  \frac{\left(\prod_{i=1}^4\hat{p}_i \right)^2}
{\hat{p}_\mu\hat{p}_\nu},
\end{eqnarray}
From the gauge invariance condition
$\sum_\mu\partial'_\mu A'_{\mu\nu}(s-s')=0$, we get
$\sum_\mu \hat{p}_\mu A'_{\mu\nu}(p)=0$.
The inverse of  $A_{\mu\nu}^{\prime \rm GF}(p)$ 
is as follows:
\begin{eqnarray}
A_{\mu\nu}^{\prime {\rm GF} -1}(p)&=&
D_{\mu\nu}(p)
+\frac{1}{\lambda}\frac{\hat{p}_\mu \hat{p}_\nu}{(\sum_i \hat{p}_i^2)^2}
e^{i(p_\mu-p_\nu)/2},
\end{eqnarray}
where 
$D_{\mu\nu}(p)$ is the $\lambda$ independent part of 
$A_{\mu\nu}^{\prime {\rm GF} -1}(p))$.
To derive the explicit form of $D_{\mu\nu}(p)$ is not so easy. However,
evaluating the determinant and cofactors of the matrix
$A_{\mu\nu}^{\prime {\rm GF}}(p)$, we can get 
\begin{eqnarray}
D_{\mu\nu}&=&
\frac{3}
  { \left\{
      \left( \sum_{a}A'_{aa} \right)^3
     -3\left( \sum_{a}A'_{aa} \right)\left( \sum_{ab}A'_{ab}A'_{ba} \right)
     +2\left( \sum_{abc}A'_{ab}A'_{bc}A'_{ca} \right)
    \right\} }
\nonumber \\
&&
\times
  \left[
    \left\{
      \left( \sum_{a}A'_{aa} \right)^2 - \left( \sum_{ab}A'_{ab}A'_{ba} \right)
    \right\}\delta_{\mu\nu}
   -2\left( \sum_{a}A'_{aa} \right)A'_{\mu\nu}
   +2\sum_{a}A'_{\mu a}A'_{a \nu}
  \right.
\nonumber \\
&&\qquad
  \left.
   -\left\{
      \left( \sum_{a}A'_{aa} \right)^2 - \left( \sum_{ab}A'_{ab}A'_{ba} \right)
    \right\}\frac{ \hat{p}_\mu\hat{p}_\nu }{ \sum_{a}\hat{p}_a^2 }
  \right]e^{i(p_\mu-p_\nu)/2}.
\end{eqnarray}
Since $\sum_\mu\partial'_\mu K_\mu(bs^{(n)})=
\sum_\mu\partial'_\mu B_\mu(bs^{(n)})=0$,
 Eq.(\ref{opwil:1}) is independent of the gauge parameter $\lambda$.

Now let us evaluate the spectrum of the  monopole current 
$K_\mu(s^{(n)})$ on the coarse lattice. Define an operator with definite 
spatial momentum $\vec{p}$:
\begin{eqnarray*}
K_i(\vec{p})_{x_4}=\int_{-\pi}^{+\pi}\frac{dp_4}{2\pi}
K_i(\vec{p},p_4)e^{ip_4x_4}.
\end{eqnarray*}
Then the correlation function is written as  
\begin{eqnarray}
\langle K_i(\vec{x},0)K_i(\vec{p})_{x_4} \rangle
&=&\int_{-\pi}^{+\pi}\frac{d^4k}{(2\pi)^4}\int_{-\pi}^{+\pi}\frac{dp_4}{2\pi}
e^{i\vec{k}\cdot\vec{x}}e^{ip_4x_4}
\langle K_i(\vec{k},k_4)K_i(\vec{p},p_4) \rangle.\label{k-k}
\end{eqnarray}
Since the monopole action on the b-lattice is written as 
$\sum_{s,s'\atop{\mu,\nu}}K_{\mu}(s)D_{\mu\nu}(s-s')K_{\nu}(s')$,
we see that the spectrum is essentially fixed by the gauge invariant 
part of the inverse of 
$D_{\mu\nu}(s-s')$. In (\ref{k-k}), it is
$\delta^4(k+p)A'_{ii}(\vec{p},p_4)$, where
\begin{eqnarray*}
A'_{ii}(\vec{p},p_4)&=&  \frac{\left(\prod_{i=1}^4\hat{p}_i \right)^2}
       {\hat{p}_i\hat{p}_i}
\prod_{i=1}^4\!\!\left( \sum_{l_i=-\infty}^{\infty} \right)
\!\!
\Biggl\{
  \!D_0^{-1}(p+2\pi l)
  \Biggl[
    \delta_{ii}\!-\frac{(p+2\pi l)_i(p+2\pi l)_i}{\sum_i(p
+2\pi l)_i^2}
  \Biggr]\\
&&\hspace{2cm}
  \frac{(p+2\pi l)_i(p+2\pi l)_i}{\prod_i(p+2\pi l)_i^2}
\Biggr\}.
\end{eqnarray*}
The integral
\begin{eqnarray*}
\int_{-\pi}^{+\pi}\frac{dp_4}{2\pi}
e^{i\vec{k}\cdot\vec{x}}e^{ip_4x_4}A'_{ii}(\vec{p},p_4)
\end{eqnarray*}
can be performed when we change $p_4+2\pi l_4 \to p_4$ and 
$\sum_{l_4}\int_{-\pi+2\pi l_4}^{\pi+2\pi l_4} \to
\int_{-\infty}^{\infty}$. Here 
\begin{eqnarray*}
D_0^{-1}(p)=
\kappa\left( \frac{m_1^2}{p^2+m_1^2}-\frac{m_2^2}{p^2+m_2^2} \right),
\end{eqnarray*}
where $\kappa$, $m_1$ and $m_2$ 
are expressed by the original couplings in Eq.(\ref{eqn.S}) as   
$\kappa(m_1^2-m_2^2)=\gamma^{-1}$, $m_1^2+m_2^2=\alpha/\gamma$ and
$m_1^2m_2^2=\beta/\gamma$.
Hence the $p_4$ integral gives us 
\begin{eqnarray*}
e^{-E_i(\vec{p}+2\pi\vec{l})x_4}, 
\end{eqnarray*}
where $E_i(\vec{p}+2\pi\vec{l})^2=-p_4^2=(\vec{p}+2\pi\vec{l})^2+m_i^2$.
The spectrum is identical with the one of the continuum.

The above monopole action can be transformed exactly into 
that of the string model
\cite{misha93,maxim98}. 
When use is made of the Poisson sum rule, we 
 write the monopole part of Eq.(\ref{opwil:1}) as
\begin{eqnarray}
&&
  \sum_{K_{\mu}=-\infty \atop{\partial^{\prime}_{\mu}K_{\mu}=0}}^{\infty}
  \exp
  \Biggl\{
    - \sum_{s,s'\atop{\mu,\nu}}
    K_{\mu}(s)D_{\mu\nu}(s-s')K_{\nu}(s') 
    + 2 \pi i \sum_{s,s'\atop{\mu,\nu}}
    B_{\mu}(s)D_{\mu\nu}(s-s')K_{\nu}(s')
 \Biggr\}
\nonumber\\
&&\qquad
=
  \int_{-\infty}^{+\infty}\!\!\!\!{\cal D}F_{\mu}(s) 
  \int_{-\pi}^{+\pi}\!\!\!\!{\cal D}\phi(s)
  \!\!\!\!\sum_{K_{\mu}(s)=-\infty}^{\infty}\!\!\!\!
  \exp
  \Bigg\{
    -\sum_{s,s'\atop{\mu,\nu}}
    F_{\mu}(s)D_{\mu\nu}(s-s')F_{\nu}(s')
\nonumber \\
&&\qquad\qquad\qquad\qquad\qquad
    + i \sum_{s,\mu}
    F_{\mu}(s)
    \Big[
      \partial_{\mu}\phi(s)
      +2\pi
      \!\!\sum_{s',\mu,\nu}\!\!
      D_{\mu\nu}(s-s')B_{\nu}(s')
    \Big]
  \Bigg\} ,\nonumber \\
&=&
  \int_{-\pi}^{+\pi}\!\!\!\!{\cal D}\phi(s)
  \!\!\!\!\sum_{l_{\mu}(s)=-\infty}^{\infty}\!\!\!\!
  \exp
  \Bigg\{
    -\frac{1}{4}\sum_{s,s'\atop{\mu,\nu}}
    \Big[
      \partial_{\mu}\phi(s) + 2\pi l_{\mu}(s)
      +2\pi
      \!\!\sum_{s_1,\mu,\alpha}\!\!
      D_{\mu\alpha}(s-s_1)B_{\alpha}(s_1)
    \Big]
\nonumber\\
&&\qquad\qquad\qquad
    D_{\mu\nu}^{-1}(s-s')
    \Big[
      \partial_{\nu}\phi(s') + 2\pi l_{\nu}(s')
      +2\pi\sum_{s_2,\nu,\beta}D_{\nu\beta}(s'-s_2)B_{\beta}(s_2)
    \Big]
  \Bigg\}.\label{st:1}
\end{eqnarray}
Performing BKT transformation and Hodge decomposition, we obtain
\begin{eqnarray}
l_{\mu}(s)
&=&
  s_{\mu}(s) + \partial_{\mu}r(s)
\nonumber \\
&=&
  \partial_{\mu}
  \Big\{
    -\sum_{s'} \Delta^{-1}_{L\;s,s'}
    \partial'_{\nu} s_{\nu}(s') + r_{\mu}(s')
  \Big\}
  +\sum_{s'} \partial'_{\nu}\Delta^{-1}_{L\;s,s'}\sigma_{\nu\mu}(s'),
\end{eqnarray} 
where $\sigma_{\nu\mu}(s) \equiv \partial_{[ \mu}s_{\nu ]}$ is the
closed string variable satisfying the conservation rule
\begin{eqnarray}
\partial_{[\alpha}\sigma_{\mu\nu]}=
  \partial_{\alpha}\sigma_{\mu\nu}+\partial_{\mu}\sigma_{\nu\alpha}
  +\partial_{\nu}\sigma_{\alpha\mu}=0.\label{conv}
\end{eqnarray}
The compact field $\phi(s)$ is absorbed into a non-compact 
field $\phi_{NC}(s)$. 
Integrating out the auxiliary non-compact field,
we see 
\begin{eqnarray}
(\ref{st:1})
&=&
  \!\!\!\!
  \sum_{\sigma_{\mu\nu}(s)=-\infty
        \atop{\partial_{[\alpha}\sigma_{\mu\nu]}(s)=0}}^{\infty}
  \!\!\!\!
  \exp
  \Bigg\{
    -\pi^2\sum_{ s,s'\atop{ \mu\neq\alpha\atop{ \nu\neq\beta } } }
      \sigma_{\mu\alpha}(s)\partial_{\alpha}\partial_{\beta}'
      D_{\mu\nu}^{-1}(s-s_1)\Delta_L^{-2}(s_1-s')\sigma_{\nu\beta}(s')
\nonumber \\
&&\qquad\qquad
  -2\pi^2\sum_{s,s'\atop{\mu,\nu}}
    \sigma_{\mu\nu}(s)\partial_{\mu}
    \Delta_L^{-1}(s-s')B_{\nu}(s') 
  - \pi^2\sum_{s,s'\atop{\mu,\nu}}
    B_{\mu}(s)D_{\mu\nu}(s-s')B_{\nu}(s')
  \Bigg\}.
\label{opwil:4}
\end{eqnarray}
The term independent of the string variable exactly cancels the second
classical term of Eq.(\ref{opwil:1}). We find finally
\begin{eqnarray}
\langle W_m({\cal C}) \rangle
&=&
  \frac{1}{Z}
  \exp
  \Bigg\{
    -\pi^2 \int_{-\infty}^{\infty}\!\!\!\!d^4xd^4y
    \sum_{\mu}N_{\mu}(x)D_0^{-1}(x-y)N_{\mu}(y)
  \Bigg\}\nonumber\\
&&
\times
  \!\!\!\!
  \sum_{\sigma_{\mu\nu}(s)=-\infty
        \atop{\partial_{[\alpha}\sigma_{\mu\nu]}(s)=0}}^{\infty}
  \!\!\!\!
  \exp
  \Bigg\{
    -\pi^2\sum_{ s,s'\atop{ \mu\neq\alpha\atop{ \nu\neq\beta } } }
    \sigma_{\mu\alpha}(s)\partial_{\alpha}\partial_{\beta}'
    D_{\mu\nu}^{-1}(s-s_1)\Delta_L^{-2}(s_1-s')\sigma_{\nu\beta}(s')
\nonumber \\
&&\qquad\qquad\qquad
    -2\pi^2\sum_{s,s'\atop{\mu,\nu}}\sigma_{\mu\nu}(s)\partial_{\mu}
    \Delta_L^{-1}(s-s')B_{\nu}(s') 
  \Bigg\}.
\label{opwil:5}
\end{eqnarray}

It is very interesting that we can evaluate  analytically 
the potential between the static electric charges when 
the monopole action on the dual lattice 
is in the weak coupling region for large $b$ as 
realized in the infrared region of pure $SU(2)$ and $SU(3)$ QCD. 
Then the string
model on the original lattice is in the strong coupling region. 
As shown later explicitly, 
the potential between the static electric charges is then evaluated
mainly  by the first 
classical part of Eq.(\ref{opwil:5}) alone. Hence let us evaluate first
the classical part.
Since the classical part is written in the continuum form, the 
continuum rotational invariance for any b lattice site 
is trivial.

The plaquette variable $S_{\alpha\beta}$ in Eq.(\ref{eqn.N}) for the 
static potential $V(Ib,0,0)$ is expressed by
\begin{eqnarray}
S_{\alpha\beta}(z)
&=&
  \delta_{\alpha 1}\delta_{\beta 4}\delta(z_{2})\delta(z_{3})
  \theta(z_{1})\theta(Ib-z_{1})
  \theta(z_{4})\theta(Tb-z_{4}).
\end{eqnarray}
Also the variable $S_{\alpha\beta}$ for the static potential
$V(Ib,Ib,0)$ is given by
\begin{eqnarray}
S_{\alpha\beta}(z)
&=&
  \Bigl(
    \delta_{\alpha 1}\delta_{\beta 4}+\delta_{\alpha 2}\delta_{\beta 4}
  \Bigr)
  \delta(z_{3})\theta(z_{4})\theta(Tb-z_{4})
\nonumber \\
&&
\times
  \theta(z_{1})\theta(Ib-z_{1})
  \theta(z_{2})\theta(Ib-z_{2})
  \delta(z_{1}-z_{2}).
\end{eqnarray}

Let us evaluate $V(Ib,0,0)$ as an example.
The Fourier transform of $S_{\alpha\beta}(z)$ in this case is 
\begin{eqnarray}
S_{\alpha\beta}(p)
&=&
\Bigl(
  \delta_{\alpha 4}\delta_{\beta 1}-\delta_{\alpha 1}\delta_{\beta 4}
\Bigr)
\int_{0}^{Ib}\!\!\!\!dz_{1}e^{-ip_{1}z_{1}}
\int_{0}^{Tb}\!\!\!\!dz_{4}e^{-ip_{4}z_{4}},
\nonumber\\
&=&
\Bigl(
  \delta_{\alpha 4}\delta_{\beta 1}-\delta_{\alpha 1}\delta_{\beta 4}
\Bigr)
\left(\frac{2}{p_{1}}\right)e^{-i\frac{p_{1}bI}{2}}\sin(\frac{p_{1}Ib}{2})
\left(\frac{2}{p_{4}}\right)e^{-i\frac{p_{4}bT}{2}}\sin(\frac{p_{4}Tb}{2}).
\label{Sk}
\end{eqnarray}
Since we study large $T$ and large $b$ behaviors, we use the
following formula:
\begin{eqnarray}
\lim_{T\rightarrow\infty}
\left(\frac{\sin \alpha T}{\alpha}\right)^2
&=& \pi T \delta(\alpha).
\label{pot:1}
\end{eqnarray}
We get 
\begin{eqnarray}
{\langle W(Ib,0,0,Tb) \rangle \atop{} }
{\longrightarrow\atop{T\rightarrow\infty\atop{b\rightarrow\infty}}}
{\displaystyle
\exp\left\{
   -\pi^2 (TIb^2)\int\frac{d^2p}{(2\pi)^2}
   \left[\frac{1}{\Delta D_0}\right](0,p_{2},p_{3},0)    
   \right\}
\atop{}
}.
\label{wil-cl-a}
\end{eqnarray}
Similarly we can evaluate $\langle W(Ib,Ib,0,Tb) \rangle$ from the
classical term.
The static potentials $V(Ib,0,0)$ and $V(Ib,Ib,0)$ can be written as
\begin{eqnarray}
V(Ib,0,0) &=&
  \pi^2 (Ib) \int\frac{d^2p}{(2\pi)^2}
  \left[
    \frac{1}{\Delta D_0}
  \right](0,p_2,p_3,0),
\nonumber \\
&=& \frac{\pi\kappa Ib}{2} \ln\frac{m_1}{m_2},\\
V(Ib,Ib,0) &=& \frac{\sqrt{2}\pi\kappa Ib}{2} \ln\frac{m_1}{m_2}. 
\end{eqnarray}
The potentials from the classical part take only the linear form and the 
rotational invariance is recovered completely even for the 
nearest $I=1$ sites. 
The string tension from the classical part is 
evaluated as 
\begin{eqnarray}
\sigma_{cl}=\frac{\pi\kappa}{2} \ln\frac{m_1}{m_2}.
\label{sigma_cl}
\end{eqnarray} 
This is consistent with the analytical results\cite{suzu89} in Type-2
superconductor.
The two constants $m_1$ and $m_2$ may be regarded as the coherence and
the penetration lengths.

Next let us evaluate the quantum fluctuation coming from the interaction 
of the string variable and the classical source. 
Since we have introduced the source term corresponding to the Wilson
loop on the fine $a$ lattice, the recovery of the rotational invariance 
of the static potential is naturally expected also for the quantum
fluctuation. Hence here we evaluate the quantum fluctuation for the flat 
Wilson loop $W(Ib,0,0,Tb)$. Then it is to be emphasized that 
the same static potential for the flat Wilson loop can be
obtained for $I,T \to \infty$ when we consider the naive Wilson loop
operator on the course $b$ lattice instead of  that on the fine lattice 
(\ref{eqn.WC}):
\begin{eqnarray}
\tilde{W}_m({\cal C})&=&\exp\bigg( 2\pi i\sum_{s,\mu}
\tilde{N}_{\mu}(s,S^J)K_{\mu}(s) \bigg),
\label{eqn.WCb}\\
\tilde{N}_{\mu}(s,S_J)&=&\sum_{s'}
\Delta_L^{-1}(s-s')\frac{1}{2}
\epsilon_{\mu\alpha\beta\gamma}\partial_{\alpha}
\tilde{S}^J_{\beta\gamma}(s'+\hat{\mu}), \label{eqn.Nb}
\end{eqnarray}
where $\tilde{S}^J_{\beta\gamma}(s'+\hat{\mu})$ is 
a flat plaquette variable satisfying 
$\partial'_{\beta}\tilde{S}^J_{\beta\gamma}(s)=
\tilde{J}_{\gamma}(s)$ and $\tilde{J}_{\gamma}(s)$ is the 
electric current on the course lattice.

Similar arguments as above shows that 
the expectation value of the operator $\tilde{W}_m({\cal C})$ is  
expressed as follows:
\begin{eqnarray}
\langle \tilde{W}_m({\cal C}) \rangle
&=&
  \frac{1}{Z}
  \exp
  \Bigg\{
    -\pi^2 \sum_{s,s'\atop{\mu,\nu}}
    \sum_{\mu}\tilde{N}_{\mu}(s)D_{\mu\nu}^{-1}(s-s')\tilde{N}_{\nu}(s')
  \Bigg\}\nonumber\\
&&
\times
  \!\!\!\!
  \sum_{\sigma_{\mu\nu}(s)=-\infty
        \atop{\partial_{[\alpha}\sigma_{\mu\nu]}(s)=0}}^{\infty}
  \!\!\!\!
  \exp
  \Bigg\{
    -\pi^2\sum_{ s,s'\atop{ \mu\neq\alpha\atop{ \nu\neq\beta } } }
    \sigma_{\mu\alpha}(s)\partial_{\alpha}\partial_{\beta}'
    D_{\mu\nu}^{-1}(s-s_1)\Delta_L^{-2}(s_1-s')\sigma_{\nu\beta}(s')
\nonumber \\
&&\qquad\qquad\qquad
    -2\pi^2\sum_{s,s'\atop{\mu,\nu}}\sigma_{\mu\nu}(s)\partial_{\mu}
    \Delta_L^{-1}(s-s_1)D_{\nu\alpha}(s_1-s')\tilde{N}_{\alpha}(s') 
  \Bigg\}.
\label{opwil:6}
\end{eqnarray}
Here we note that $D_{\mu\nu}^{-1}(s-s')$ is given by
Eq.(\ref{amunupr}) and that the Fourier transform 
$\tilde{S}_{\beta\gamma}(k)$ of
$\tilde{S}^J_{\beta\gamma}(s)$ is similar to Eq.(\ref{Sk}) with the
momentum defined on the course lattice.
 Hence the first classical part
becomes 
\begin{eqnarray}
&&\pi^2\sum_{s,s'\atop{\mu,\nu}}N_{\mu}(s)D_{\mu\nu}^{-1}(s-s')N_{\nu}(s')
\nonumber \\
&=&4\pi^2\epsilon_{\mu\alpha 14}\epsilon_{\nu\beta 14}\int_{-\pi}^{\pi}
\frac{d^4k}{(2\pi)^4}\Delta_L^{-2}(k)
\sin\left( \frac{k_\alpha}{2} \right) \sin\left( \frac{k_\beta}{2} \right)
e^{i(\frac{k_\alpha}{2}-\frac{k_\beta}{2})}\nonumber \\
&&\times\ \ \  \tilde{S}_{14}(k)\tilde{S}_{14}(-k)
\sum_l \left(
  \delta_{\mu\nu}-\frac{(k+2\pi l)_\mu(k+2\pi l)_\nu}{(k+2\pi l)^2}
\right)D_0^{-1}(k+2\pi l)\nonumber \\
&&\times\ \ \  \Pi_{i\neq\mu \atop{j\neq\nu}}\pi_i(k+2\pi l)\pi_j^*(k+2\pi l),
\label{wil-cl-b}
\end{eqnarray}
where 
\begin{eqnarray}
\pi_i(k)=\frac{\sin (k_i/2)}{k_i/2}e^{-ik_i/2}.
\end{eqnarray}
Changing the integral variable as $k+2\pi l\to k$, we can absorb the
summation with respect to $l$ using the integral
over the infinite momentum range. When we use
Eq.(\ref{pot:1}) for large $I$ and $T$, we find the classical part   
(\ref{wil-cl-b}) agrees exactly with (\ref{wil-cl-a}).

Similarly 
\begin{eqnarray}
&&\sum_{s'}D_{\mu\nu}^{-1}(s-s')N_{\nu}(s') \nonumber \\
&=&\int_{-\pi}^{\pi}
\frac{d^4k}{(2\pi)^4}
\sum_l \left(
  \delta_{\mu\nu}-\frac{(k+2\pi l)_\mu(k+2\pi l)_\nu}{(k+2\pi l)^2}
\right)D_0^{-1}(k+2\pi l)\nonumber \\
&\times&\ \ \Pi_{i\neq\mu \atop{j\neq\nu}}
\pi_i^*(k+2\pi l)\pi_j(k+2\pi l)e^{ik_\mu}
2i\epsilon_{\nu\alpha 14}\Delta_L^{-1}(k)
\sin\left( \frac{k_\alpha}{2} \right)
e^{i\frac{k_\alpha}{2}}\frac{\tilde{\pi}_1(k)\tilde{\pi}_4(k)}
{\pi_1(k)\pi_4(k)}e^{iks},
\label{wil-q-b}
\end{eqnarray}
where 
\begin{eqnarray}
\tilde{\pi}_1(k)&=&\frac{\sin(k_1I/2)}{k_1/2}e^{-ik_1I/2},
\ \ \ \ \tilde{\pi}_4(k)=\frac{\sin(k_4T/2)}{k_4/2}e^{-ik_4T/2}.
\end{eqnarray}
When use is made of 
\begin{eqnarray}
\sin\left(\frac{k_1I}{2}\right)\sin\left(\frac{k_1}{2}\right)
=\sin^2\left(\frac{k_1(I+1)}{4}\right)
-\sin^2\left(\frac{k_1(I-1)}{4}\right),
\end{eqnarray}
we get for large $I$ and $T$
\begin{eqnarray}
\pi_1^{*}(k)\tilde{\pi}_1(k)\pi_4^{*}(k)\tilde{\pi}_4(k)
=(2\pi)^2\delta(k_1)\delta(k_4)e^{i(1-I)k_1/2}e^{i(1-T)k_4/2}.
\end{eqnarray}
We find for large $I$ and $T$ that 
$\sum_{s'}D_{\mu\nu}^{-1}(s-s')N_{\nu}(s')$ is equivalent to 
$B_\mu(s)$ in (\ref{opwil:9}).

Hence the expectation value of the naive Wilson loop
$\langle \tilde{W}_m({\cal C}) \rangle $ in (\ref{opwil:6}) coincides with 
that of the perfect operator in (\ref{opwil:5}).
Now we introduce the dual string variable ${}^*\sigma$ as follows:
\begin{eqnarray}
\sigma_{\mu\nu}(s)
&=&\frac{1}{2}\epsilon_{\mu\nu\alpha\beta}{}^*\sigma_{\alpha\beta}
(s+\hat{\alpha}+\hat{\beta}),\nonumber \\
{}^*\bar{\sigma}_{\alpha\beta}(s)&\equiv& {}^*\sigma_{\alpha\beta}
+S_{\alpha\beta}(s).
\end{eqnarray}
Then the expectation value (\ref{opwil:6}) can be expressed simply as 
\begin{eqnarray}
\langle \tilde{W}_m({\cal C}) \rangle
&=&
  \frac{\sum_{{}^*\bar{\sigma}_{\mu\nu}(s)=-\infty}^{\infty}
\Pi_{s \atop{\mu,\nu}}\delta(\partial_\mu'{}^*\bar{\sigma}_{\mu\nu}(s)
-\tilde{J}_\nu (s))\exp\{-S({}^*\bar{\sigma})\}}
{\sum_{{}^*\bar{\sigma}_{\mu\nu}(s)=-\infty}^{\infty}
\Pi_{s \atop{\mu,\nu}}\delta(\partial_\mu'{}^*\bar{\sigma}_{\mu\nu}(s))
\exp\{-S({}^*\bar{\sigma})\}},
\end{eqnarray}
where 
\begin{eqnarray}
S({}^*\bar{\sigma})&=&\pi^2\sum_{s,s'\atop{\mu,\nu}}
(\partial_{\alpha}\Delta_L^{-1}(s-s_1)\frac{1}{2}
\epsilon_{\alpha\mu\beta\gamma}{}^*\bar{\sigma}_{\beta\gamma}
(s_1+\hat{\alpha}+\hat{\mu}))D_{\mu\nu}^{-1}(s-s')\\
&&\ \ \times (\partial_{\beta}\Delta_L^{-1}(s'-s_2)\frac{1}{2}
\epsilon_{\beta\nu\eta\delta}{}^*\bar{\sigma}_{\eta\delta}
(s_2+\hat{\beta}+\hat{\nu})).
\end{eqnarray}

The strong coupling expansion can be shown in Fig.\ref{strong}. The leading
term is the same as the classical contribution in (\ref{wil-cl-b}).
The next-leading term is a house-type diagram with one $1\times 1$ cube 
attached on the flat surface. Then the open surface variable
${}^*\bar{\sigma}$ has four more  plaquettes than the leading one.
If the self coupling term between the string variables ${}^*\bar{\sigma}$
is dominant as in $SU(2)$ QCD, the next to leading term is estimated as 
$\exp(-\sigma_{cl}ITb^2-4\Pi(0)b^2)$, where $\sigma_{cl}$ is the string 
tension from the classical part (\ref{sigma_cl}) and 
$\Pi(0)$ is the self
coupling constant. Considering the entropy factor $4IT$, we get 
\begin{eqnarray}
\langle \tilde{W}_m({\cal C}) \rangle
&=& \exp\{-\sigma_{cl}ITb^2\}+4IT\exp\{-\sigma_{cl}ITb^2-4\Pi(0)b^2\}
+\cdot\cdot\cdot ,\nonumber\\
&\simeq&\exp\{-\sigma_{cl}ITb^2\}\exp\{4ITe^{-4\Pi(0)b^2}\},\nonumber\\
&\equiv&\exp\{-\sigma ITb^2\}.
\end{eqnarray}
Hence the string tension becomes 
\begin{eqnarray}
\sigma=\sigma_{cl}-\frac{4}{b^2}e^{-4\Pi(0)b^2}.
\end{eqnarray}

Applications to actual pure $SU(2)$ and $SU(3)$ QCD will be published 
elsewhere\cite{kato9910,yamagishi9910} and 
the quantum fluctuation term will be  found to be  very small there.

\acknowledgments

T.S. acknowledges the financial support from  
JSPS Grant-in Aid for Scientific Research (B) (No.10440073 and No.11695029).
 

\begin{figure}
\caption{The strong coupling expansion of the Wilson loop calculation.}
\label{strong}
\end{figure}

\newpage
\pagestyle{empty}

\epsfxsize=.8\textwidth
\begin{center}
\leavevmode
\epsfbox{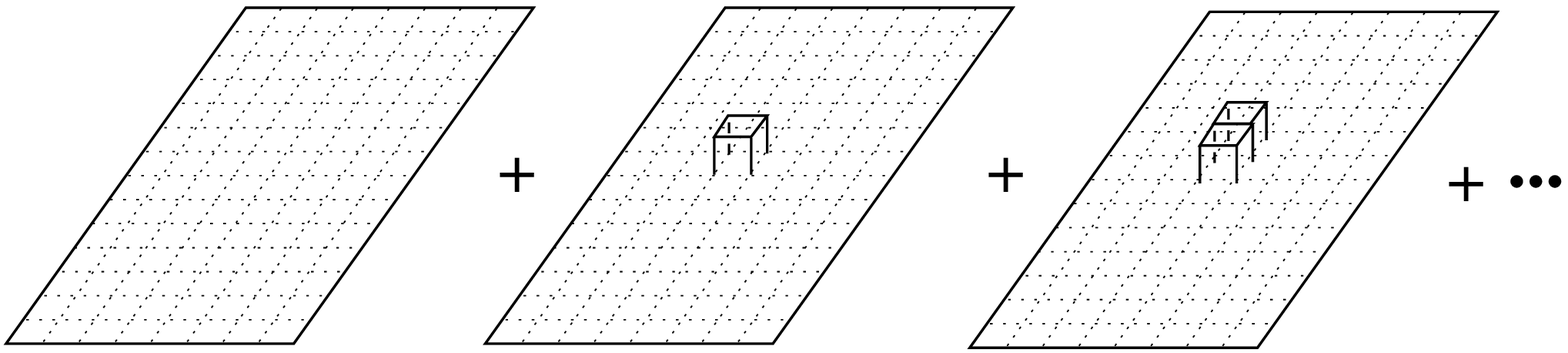}
\end{center}

\end{document}